\documentclass[aps,prb,amsmath,superscriptaddress,twocolumn,longbibliography]{revtex4-1}

\usepackage{graphicx}
\usepackage{amsmath}
\usepackage{amsfonts}
\usepackage{amssymb}
\usepackage{braket}
\usepackage{bm}
\usepackage{multirow}
\usepackage{color}
\usepackage[normalem]{ulem}
\usepackage{mathrsfs}
\usepackage{mathtools}
\usepackage{float}

\newcommand{\red}[1]{{\textcolor{red}{#1}}}

\makeatletter

\newcommand{\Rmnum}[1]{\expandafter\@slowromancap\romannumeral #1@}
\makeatother

\usepackage[breaklinks]{hyperref}
\hypersetup{colorlinks=true, linkcolor=blue, citecolor=blue, filecolor=blue, urlcolor=blue}

\AtBeginDocument{%
    \newwrite\bibnotes
    \def\bibnotesext{Notes.bib}
    \immediate\openout\bibnotes=\jobname\bibnotesext
    \immediate\write\bibnotes{@CONTROL{REVTEX41Control}}
    \immediate\write\bibnotes{@CONTROL{%
    apsrev41Control,author="08",editor="1",pages="1",title="0",year="1"}}
     \if@filesw
     \immediate\write\@auxout{\string\citation{apsrev41Control}}%
    \fi
}%

\def\i{\mathrm i}

\def\sz{\sigma^{\rm z}}
\def\sx{\sigma^{\rm x}}

\def\mz{\mu^{\rm z}}
\def\mx{\mu^{\rm x}}

\usepackage[]{lineno}

\begin{document}

\title{Experimental Realization of Classical $\mathbb{Z}_2$ Spin Liquids\\ in a Programmable Quantum Device}

\author{Shiyu Zhou}
\email{zhous@bu.edu}
\affiliation{Physics Department, Boston University, Boston, MA, 02215, USA}

\author{Dmitry Green}
\email{dmitry.green@aya.yale.edu}
\affiliation{Physics Department, Boston University, Boston, MA, 02215, USA}
\affiliation{AppliedTQC.com, ResearchPULSE LLC, New York, NY, 10065, USA}

\author{Edward D. Dahl}
\email{denny.dahl@coldquanta.com}
\affiliation{ColdQuanta, Inc., Boulder, CO 80301, USA}

\author{Claudio Chamon}
\email{chamon@bu.edu}
\affiliation{Physics Department, Boston University, Boston, MA, 02215, USA}

\date{\today}


\onecolumngrid
\begin{abstract}

  We build and probe a $\mathbb{Z}_2$ spin liquid in a programmable
  quantum device, the D-Wave DW-2000Q. Specifically, we observe the 
  classical 8-vertex and 6-vertex (spin ice) states and transitions 
  between them. To realize this state of matter, we design a 
  Hamiltonian with combinatorial gauge symmetry
  using only pairwise-qubit interactions and a transverse field, i.e.,
  interactions which are accessible in this quantum device. The
  combinatorial gauge symmetry remains exact along the full quantum
  annealing path, landing the system onto the classical 8-vertex model
  at the endpoint of the path. The output configurations from the
  device allows us to directly observe the loop structure of the
  classical model. Moreover, we deform the Hamiltonian so as to
  vary the weights of the 8 vertices and show that we can selectively
  attain the classical 6-vertex (ice) model, or drive the system
  into a ferromagnetic state. We present studies of the
  classical phase diagram of the system as function of the
  8-vertex deformations and effective temperature, which we control by
  varying the relative strengths of the programmable couplings, and we
  show that the experimental results are consistent with theoretical
  analysis. Finally, we identify additional capabilities that, if
  added to these devices, would allow us to realize
  $\mathbb{Z}_2$ quantum spin liquids on which to build topological
  qubits.

\end{abstract}
\maketitle

{\bf Introduction} --
Quantum spin liquids (QSLs) have a long history. They were first
proposed in the 1970s by Anderson~\cite{Anderson1973} as an
alternative to the spin 1/2 N\'eel antiferromagnetic state, and later
as candidates for the insulating parent state of the high-temperature
superconductors\cite{Anderson1987}. QSLs do not display magnetic
symmetry-breaking order, but instead display topological
order~\cite{Wen1990}. They are also closely related to lattice gauge
models in particle physics, also dating to the
1970s~\cite{z2_Wegner, z2_Kogut, Fradkin-Susskind}. Today, a QSL
model known as the toric code is a potential platform for topological
computing~\cite{tqc_Kitaev}. There have been many proposed materials,
but to-date gapped QSLs have not been unambiguously observed in
nature~\cite{BalentsL2010, Savary2016, RevModPhys.89.025003,
s41535-019-0151-6, Broholm2020}.  Since they are so hard to find in
materials, a more recent idea is to build a QSL synthetically out of
superconducting circuits~\cite{Ioffe1, Ioffe2, CG2020}. In parallel,
another recent idea has emerged to simulate familiar quantum phase
transitions on programmable devices~\cite{app_King, app_Harris}. In
this paper we take these ideas one step further and show that, in
principle, a programmable device can be used to emulate QSL phases so
far unreachable by other means, a step towards realizing logical
topological qubits in these same devices. While we cannot observe the
full quantum regime due to the limitations of the current device, we
do observe unmistakable signatures of the phase in its classical limit
at the endpoint of the quantum annealing protocol. In the process of
doing so, we identify additional features that a programmable device
of this sort would need in order to realize QSLs. It is a testament to
technological progress that a handful of theorists can observe and
experiment with new physics while being equipped only with remote
access to a commercial device.

From a theoretical perspective, our framework requires a new general
construct: ``Combinatorial Gauge Symmetry''~\cite{CGY}. This is an
{\it exact and non-perturbative} symmetry that stabilizes the
topological phase for a wide range of parameters.  From a practical
perspective, it enables the programming of QSLs in D-Wave because it
requires at most two-body Ising spin interactions. The vast majority
of theoretical QSL models rely on pure multi-spin interactions, which
heretofore are not attainable. The notable exception is the Kitaev
hexagonal model~\cite{hex_Kitaev}, but it requires $XX$, $YY$ and $ZZ$
interactions simultaneously, which have also been unattainable. In our
case the model that we program is the $\mathbb{Z}_2$
lattice gauge theory. The toric code~\cite{tqc_Kitaev}, which is a
central model for topological quantum computing, is a special limit of
the $\mathbb{Z}_2$ model.

The D-Wave DW-2000Q quantum annealer comprises a superconducting
circuit which implements programmable $ZZ$-couplings between pairs of
spins (qubits): $J_{ij}\sz_i\sz_j$. Programmability means that the
$J_{ij}$ coefficients can be specified, although the available
couplings are constrained by the connectivity of the device. A
transverse field $\Gamma$ is applied to each spin, coupling to the
x-component of the spins, $\sx_i$, inducing spin flips. The machine
can be operated with the forward annealing protocol described in the
device documentation~\cite{D-WaveDoc}, in which all active
spins are initialized in the ground state of a Hamiltonian with
a high transverse field $\Gamma$ that is gradually decreased to zero,
while the $J_{ij}$ couplings are slowly increased from zero to their
specified final values. At the endpoint of the protocol the system
should land in a ground state of the classical Hamiltonian
associated with the $J_{ij}$'s if a dynamical obstruction (such as a
glass transition) is not encountered.

\begin{figure*}[!tbh]
\centering
\includegraphics[width=.9\textwidth]{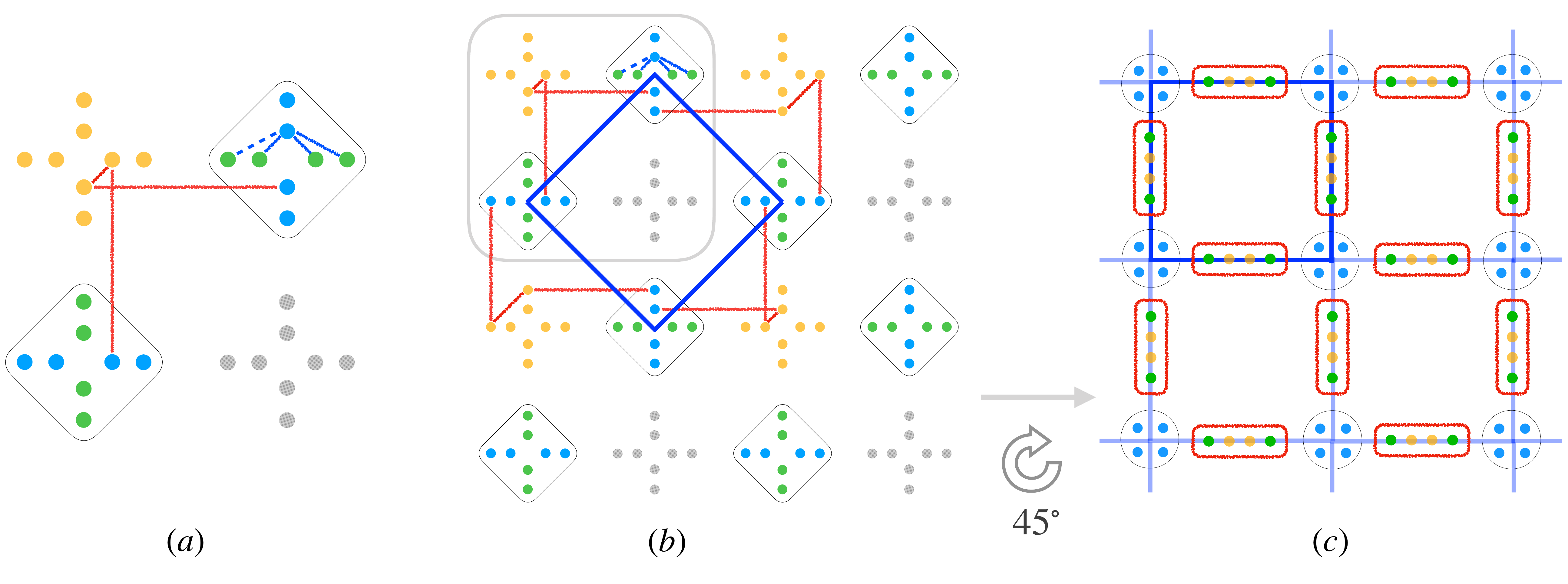}
\caption{(a) Embedding details for two adjacent star operators in
  Eq.~\eqref{eq:Hamiltonian}. Each boxed cell represents one star
  operator consisting of four gauge spins $\sigma_i$ (green dots) and
  four matter spins $\mu_a$ (blue dots). In a star, gauge spins only
  couple to matter spins according to the $W$ matrix in
  Eq.~\eqref{eq:W}, modulated by the strength $J$ (blue lines). Only
  one set of $J$ connections is shown for clarity. Note that one of
  the couplings (shown by a dashed blue line) is different from the
  other three (solid blue lines) and corresponds to the diagonal
  entries in the $W$ matrix. Ghost spins (orange dots) are the
  necessary bridges to connect adjacent stars together. Each
  $\sigma_i$ is copied to two ghost spins in the nearby unit cell with
  strong ferromagnetic coupling $K$ (shown by a red line). There are
  unused spins in this embedding (grey dots). (b) The zoomed out view
  of the embedding, showing more stars connected through ghost
  spins. A plaquette of the effective lattice is indicated by thick
  blue lines. (c) The construction rotated through $45^\circ$, is
  equivalent to a lattice model where the effective gauge spins
  $\sigma_i$ are represented by the four strongly coupled qubits that
  act as one. These four spins (two blue and two orange) reside on the
  bonds. The matter spins $\mu_a$ reside on the vertices (blue dots)
  of the effective lattice.}
\label{fig:embedding} 
\end{figure*}

In this work, we program the $J_{ij}$ coefficients so that the
spectrum represents a system with exact $\mathbb{Z}_2$ gauge symmetry.
We observe and study the Hamiltonian in the classical regime at the
endpoint of the annealing path, and map out its phases, which include
both 8- and 6-vertex models. The 8-vertex model is the classical
version of the toric code, while the 6-vertex phase is also known as
planar spin ice ~\cite{Lieb1967}. Both planar~\cite{Mengotti2010} and
three-dimensional~\cite{Ramirez1999} spin ice have been observed; in
our construction, we observe planar ice as a special case. In
contrast, to our knowledge, neither the classical nor quantum 8-vertex
models have been observed. We synthesize them here and probe the
classical version.

{\bf Model and Embedding} --
The model is based on a square lattice, where we place ``gauge" qubits
on each link and four ``matter" qubits at each vertex. The embedding
of this geometry within the available D-Wave architecture is shown in
Fig.~\ref{fig:embedding}. First, each gauge qubit is coupled to its
neighboring matter qubits with strength proportional to $J$ within the
unit (Chimera) cell of the device. Second, we utilize four
strongly-coupled qubits, with strength $K$, to effectively act as one
gauge qubit, which is required because of the constrained coordination
number of the Chimera architecture~\cite{D-WaveDoc}. To flip an
effective gauge qubit, a transverse field must flip a total of four
device qubits, so the effective transverse field on the gauge qubits
is of the order $\widetilde\Gamma\sim \Gamma^4/K^3$.

The embedding as described above is a realization of the following
Hamiltonian
\begin{align}
H=-\sum_s
\left[
  J\sum_{\substack{a\in s\\i\in s}} \,W^{}_{ai}\;\sz_i\,\mz_a
  +\Gamma \,\sum_{a\in s} \mx_a
  \right]-\widetilde\Gamma \sum_i \sx_i
\;,
\label{eq:Hamiltonian}
\end{align}
where $\sigma_i$ and $\mu_a$ are Pauli matrices representing the gauge
and matter spins, respectively, around each vertex $s$. The $4\times
4$ interaction matrix,
\begin{align}
  W =
  \begin{pmatrix}
    -1+\eta&1&1&1\\
    1&-1+\eta&1&1\\
    1&1&-1+\eta&1\\
    1&1&1&-1+\eta
  \end{pmatrix}\;,
  \label{eq:W}
\end{align}
encodes the ferromagnetic/anti-ferromagnetic interactions between
gauge and matter spins that are programmed into the device.

\begin{figure}[!h]
\centering
\includegraphics[width=0.45\textwidth]{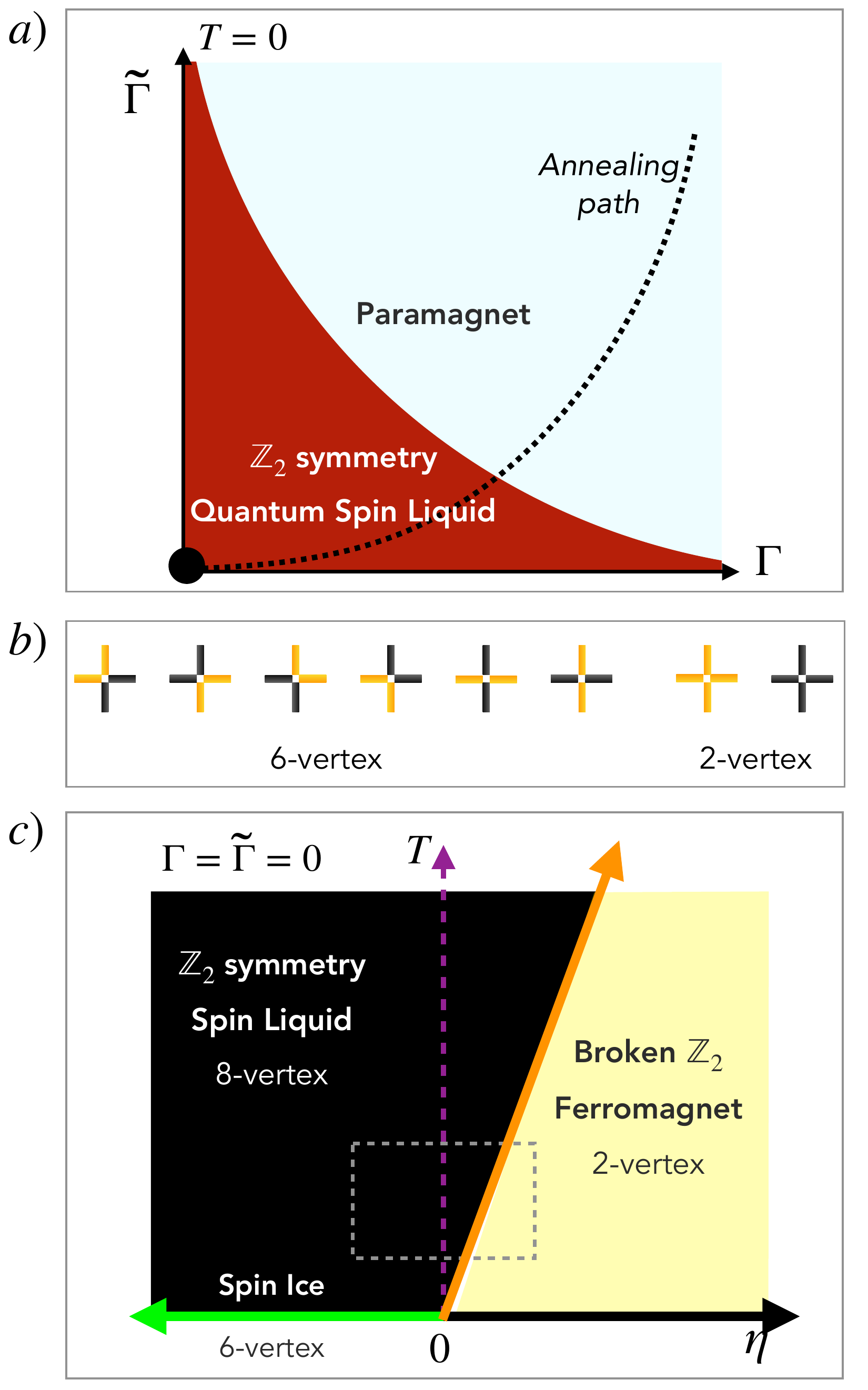}
\caption{(a) Schematic phase diagram of the Hamiltonian in
  Eq.~(\ref{eq:Hamiltonian}) for $\eta=0$. The annealing path goes
  from the paramagnetic to the quantum spin liquid with $\mathbb{Z}_2$
  symmetry. The Hamiltonian obeys this symmetry along the entire
  annealing path. The system allows only one transverse field, however
  there are two effective ones, $\Gamma$ and $\widetilde \Gamma$, as a
  result of the embedding as described in the text. Measurements are
  taken at the classical limit where $\Gamma = \widetilde
  \Gamma=0$. (b) Vertex configurations for the gauge spins in the
  ground state. The 6-vertex and 2-vertex groupings are separated by
  an energy gap $\delta=4\eta$. (c) Phase diagram for the classical
  limit ($\Gamma = \widetilde \Gamma=0$ ) of the $8-$vertex model as a
  function of $\eta$ and $T$. The dotted box depicts a window of
  $\eta$ and $T$ parameters where experimental measurements are
  accessible.}
\label{fig:annealing} 
\end{figure}

\begin{figure*}[!t]
\centering
\includegraphics[width=0.9\textwidth]{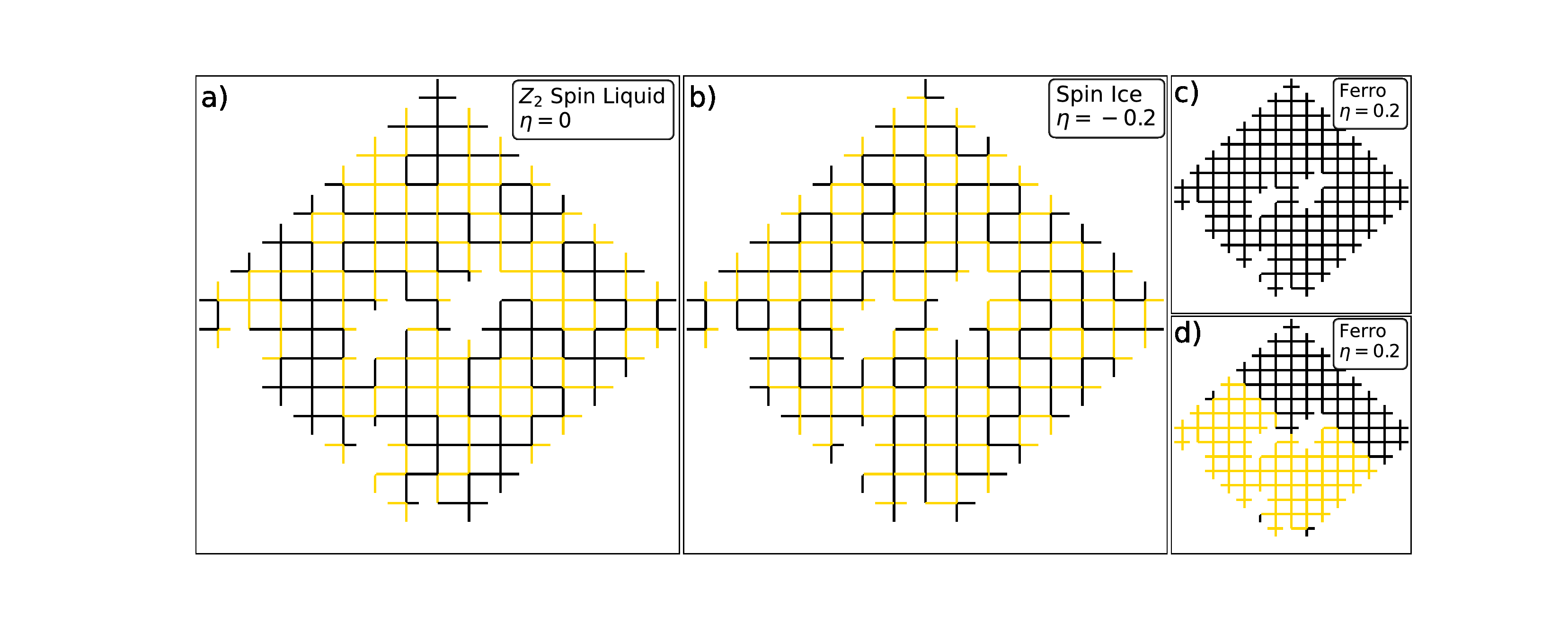}
\caption{Four typical configurations of final states returned by the
  device at the endpoint of the annealing cycle, obtained at base
  temperature ($\tau=1$) for different values of the parameter
  $\eta$. The color of the bonds represent the state of the gauge spin
  [associated to that of the four strongly coupled qubits on the
    bonds, see Fig.~\ref{fig:embedding}(c)]: up gauge spins are
  colored in gold, and down spins are colored in black. The matter
  spins, which are tethered to the gauge spins at low energies, are
  omitted in the picture. The blank areas indicate punctures on the
  lattice due to the inactive qubits in the machine. (a) at $\eta =
  0$, a $\mathbb Z_2$ spin liquid state is observed where closed loops
  or connected strings of either spin up or down are formed, and can
  intersect; (b) at $\eta = -0.2$, a spin ice state (or $6$-vertex
  model) is obtained where similar loops and strings are formed, but
  do not intersect. (c) at $\eta = 0.2$, system fully magnetizes; and
  (d) at $\eta = 0.2$, two ferromagnetic domains with opposite
  magnetization are separated by a domain wall. }
\label{fig:config} 
\end{figure*}

\begin{figure}[!tbh]
\centering
\includegraphics[width=0.4625\textwidth]{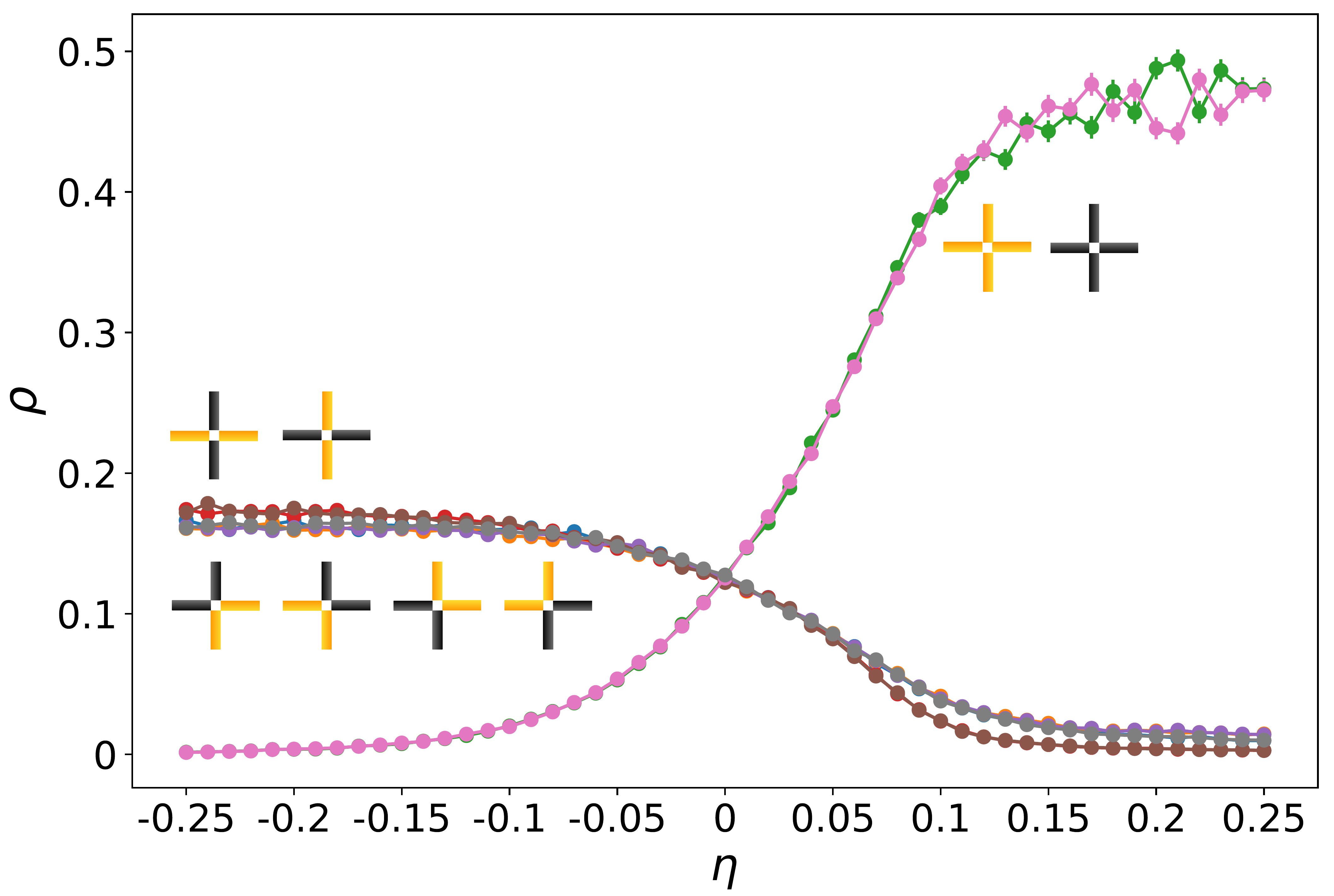}
\caption{Densities of all 8-vertex types for $\eta$ varying from
  $-0.25$ to $0.25$, at base temperature ($\tau=1$). The data clusters
  according to three different families of stars, indicated next to
  the curves. At $\eta=0$ all 8-vertex types appear with equal
  densities, consistent with the presence of the combinatorial gauge
  symmetry. The 6-vertex types are favored for $\eta<0$, and the
  2-vertex types for $\eta>0$.}
\label{fig:8-types}
\end{figure}

At the special point $\eta=0$, $W$ is a $4\times 4$ Hadamard matrix,
which has the property that $W^\top W=4\, {\mathbb I}$ and every entry
is equal to $\pm1$. This point is characterized by a local symmetry
which is generated by a group of monomial matrices that represent
flips of single spins as well as permutations among matter spins on
each vertex. This local combinatorial symmetry ensures that the
Hamiltonian obeys a $\mathbb Z_2$ gauge symmetry for \textit{any value
  of} $J$, $\Gamma$ and $\widetilde\Gamma$.~\cite{CGY}

At zero-temperature, for $\Gamma$ and/or $\widetilde\Gamma$ small, the
model is in its topological phase ~\cite{CGY}, as schematically
depicted in Fig.~\ref{fig:annealing}(a). The energy gap of this QSL
phase is on the order of $10^{-3}-10^{-2}J$~~\cite{Kaishin}, and the
temperature of the device renders the gap unobservable. (We note that
dynamics in this regime may still reveal coherent propagation of
quasiparticles, see Ref.~\onlinecite{Hart-etal} for a theoretical
discussion.) While the QSL is inaccessible in the current devices, the
annealing cycle in D-Wave, designed to adiabatically transform the
system from a region of finite to zero transverse field, lands the
system at a classical $\mathbb Z_2$ spin liquid state at
$\Gamma=\widetilde\Gamma=0$. Because the Hamiltonian obeys the
combinatorial $\mathbb Z_2$ gauge symmetry {\it exactly} for any value
of the couplings, the symmetry is respected throughout the entire
quantum annealing path, minimizing the number of defects at the point
of observation. The ground states at zero transverse field and
$\eta=0$ are the same as that of the 8-vertex model~\cite{Baxter}, in
which the number of up and down gauge qubits around any vertex is an
even number (even parity star), thus producing the loop structure
associated with the $\mathbb Z_2$ spin liquid state. In the
Supplementary Material~\cite{Supplementary} we show the symmetries and ground state
degeneracy explicitly.

For $\eta\neq 0$, the $W$ matrix is deformed away from the point of
combinatorial gauge symmetry, allowing us to access other
states. Non-zero $\eta$ splits the 8 vertices into groups of 2 and 6
vertices, as shown in Fig.~\ref{fig:annealing}(b), where up and down
gauge qubits are pictured as black and gold links, respectively (the
configurations of the matter qubits are tethered to those of the gauge
qubits in the ground state). The energy separation is $\delta=4J\eta$;
$\eta<0$ favors the 6-vertex model and $\eta>0$ favors the
ferromagnetic states, where all spins in a star are up or down. The
topological phase $\eta=0$ sits at a zero-temperature critical point.

At non-zero temperature, vertices with odd parity are allowed but
exponentially suppressed at low temperatures. This enables us to study
the phases of the classical 8-vertex model, whose theoretical phase
diagram is shown in Fig.~\ref{fig:annealing}(c). In our programmed
Hamiltonian we vary $\eta$ and $J$, but since the operating
temperature of the device is fixed, we are effectively varying $\eta$
and $J/T$. This corresponds to varying the relative Boltzmann weight
(fugacity) between the 6- and 2-vertex states,
$d=e^{\delta/T}=e^{4J\eta/T}$. The 8-vertex spin liquid phase is
stable in the range $0<d<3$~\cite{Ardonne2004, Fradkin2007}. In this
region there is a continuously varying imbalance between the 6- and
2-vertex types. The system transitions to an ordered ferromagnet at
$d=3$, and for $d>3$ it remains magnetized, settling to one out of the
two 2-vertex states. Strictly speaking, the $d=0$ limit exists at zero
temperature only, where $\eta<0$ and we have the 6-vertex, or spin
ice, phase. In the experiments, however, there is a small range of
temperatures where it occurs because the system size is finite.

{\bf Results} -- The experiments were run on the D-Wave DW-2000Q
machine at Los Alamos National Laboratory (LANL). The $J$ and $K$
couplings are programmed into the D-Wave's Chimera architecture
according to the embedding in Fig.~\ref{fig:embedding}. The ratio
$J/K$ is fixed to $1/2$ in the experiments, which allows the ghost
qubits to be coupled strongly while allowing sufficiently broad range
for varying $J$. The operating temperature of the device is
approximately 12 mK, but by scaling $J$ and $K$ (by the same factor)
we effectively control the ratio $T/J$. The lowest temperature in our
measurements corresponds to choosing the value
$J^{\rm API}_{\rm max}=1/2$ in the Application Programming Interface
(API) for the D-Wave DW-2000Q~\cite{D-WaveDoc}; by decreasing the
input variable to $J^{\rm API}=J^{\rm API}_{\rm max}/\tau$, where
$\tau\ge 1$, we increase the effective temperature by a factor
$\tau$. The factor can be expressed as $\tau=\alpha T/J$, where
$\alpha$ is the dimensionless ratio between the maximum physical
coupling $J_{\rm max}$ (in units of mK) to the device temperature when
$J^{\rm API}_{\rm max}=1/2$. We calibrate $\alpha$ by fitting the
spectrum of the Hamiltonian for independent stars, and find it to be
approximately 15 (see Supplementary Material~\cite{Supplementary}).

\begin{figure}[!tbh]
\centering
\includegraphics[width=0.5\textwidth]{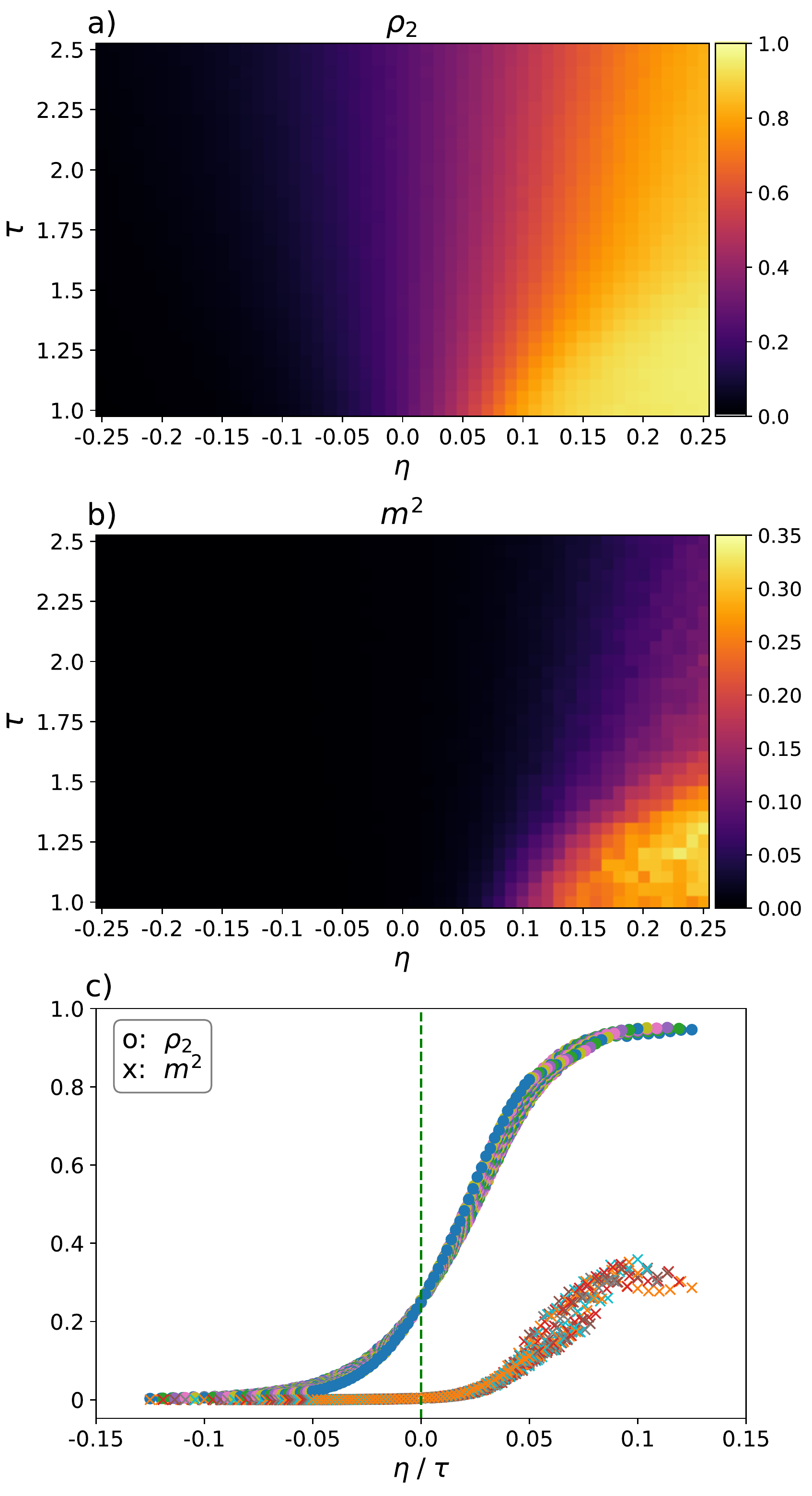}
\caption{Heat plots (a) and (b) show the observables $\rho_2$ and
  $m^2$ for $\eta$ varying from $-0.25$ to $0.25$ in intervals of
  $0.01$, and $\tau$ varying from $1$ to $2.5$ in intervals of $0.05$.
  The dependence of $\rho_2$ on the slopes $\tau/\eta$ of lines on the
  $\eta$ --- $\tau$ plane can be observed in (a). Map (b) shows that
  the magnetization vanishes in the spin liquid region (black region),
  and continuously grows past the transition to the ferromagnetic
  state (purple and yellow regions).  (c) shows $\rho_2$ and $m^2$ as
  a function of $\eta\,/\,\tau$. The collapse of the data shows that
  $\rho_2$ and $m^2$ are controlled by a single parameter, the ratio
  $\eta/\tau$. }
\label{fig:eta-f}
\end{figure}

The total annealing time in the experiments is chosen to be
$2000~\mu s$, with a pause from $20~\mu s$ to $1000~\mu s$ in
which $\Gamma$ is held at approximately half its maximum value. A
wait time of $100~\mu s$ is allowed between consecutive annealing
cycles. Each run returns $1000$ different configurations of final
states, projected onto the $z$-basis. Typical configurations for
different parameters $\eta$ and at base temperature ($\tau=1$), are
shown in Fig.~\ref{fig:config}. The total number of active stars in
the system is $N_{\rm stars}=114$ (if all qubits were active, there
would be 128 stars.)

We observe the $\mathbb{Z}_2$ spin liquid in the expected range of
parameters, with an example at $\eta=0$ in
Fig.~\ref{fig:config}(a). At each vertex, or star, the parity of the
qubits at the links is even, and each of the allowed 8-vertex
configurations occurs with equal proportion (see
Fig.~\ref{fig:8-types}). Strings are visible in
Fig.~\ref{fig:config}(a) by following a gold or black line. These
strings either form closed loops in the bulk, or are open but
terminate either at the external boundary or at the internal holes
formed by the cluster of inactive qubits. We also observe the spin
ice, or 6-vertex model, when $\eta<0$, as shown in
Fig.~\ref{fig:config}(b). The parity constraints are satisfied but the
2-vertex configurations are suppressed resulting in non-intersecting
strings. Figs.~\ref{fig:config}(c) and (d) show ferromagnetic
configurations, where the up/down symmetry is broken and there is an
imbalance between the two 2-vertex types. In (c) the fully magnetized
state is reached, while in (d) the system breaks into two domains with
opposite magnetization and separated by a domain wall.

We gather 1000 sample configurations for each of 51 values of $\eta$
equally spaced in the window $-0.25\le \eta \le 0.25$ and 31 values of
$\tau$ equally spaced in the range $1\le \tau \le 2.5$. The data
accumulated over these $51\times 31$ bins can be aggregated to yield
evidence that the 8-vertex constraint is satisfied in this window of
the $\eta$ --- $\tau$ plane. Violations of the 8-vertex constraints
take two forms in this embedding: (a) an odd parity star, or (b) a
broken gauge link where the four physical qubits representing it are
not all aligned. Out of the $51\times 31\times 1000 = 1581000$ samples
gathered, 1007751 samples (or $63.7\%$) had no defects in any of the
stars or links comprising the system, while 374217 (or $23.7\%$) had
only a single defect, 146865 (or $9.3\%$) had two defects, and 38857 (or
$2.5\%$) had three defects. We note that most defects occur
at higher temperatures, as expected, and in the ferromagnetic region
because of domain walls. At base temperature and in the region of
negative $\eta$ where the topological phases of the 8-vertex model
occur, $86.0\%$ of the samples had zero defects, $11.7\%$ had one defects,
$2.0\%$ had two defects, and $0.2\%$ had three defects. 
We note that the fraction of single stars with no defects is $99.9\%$ at 
base temperature and in the $\eta\le 0$ region. We present in the
Supplementary Material~\cite{Supplementary} a detailed study of the temperature dependence
of the defect densities, which we also use to calibrate the constant
$\alpha$ relating $\tau$ to $T/J$. The low number of defects is
evidence that the 8-vertex model is realized within the experimental
window.

The samples enable us to study several observables, such as the
densities of all 8-vertex types, shown in Fig.~\ref{fig:8-types} for
$\eta$ from $-0.25$ to $0.25$, at base temperature ($\tau=1$). At
$\eta=0$, the system is placed at the combinatorial gauge symmetry
point, and all 8-vertex types appear with equal densities. To the left
of the symmetry point, for $\eta<0$, the 6-vertex configurations are
favored for sufficiently negative $\eta$, where the densities of
2-vertex types become exponentially small and thus not observable in
the finite lattice. (We note that the densities of 2-vertex types are
strictly zero only at $T=0$.) We observe a small splitting between the
6 configurations, with slightly different densities for the two stars
that are inversion symmetric as compared to the four that are not. To
the right at $\eta>0$, the 2-vertex configurations are favored. In
this region, spontaneous magnetization is detected via the order
parameter $m^2=(\rho_{2+}-\rho_{2-})^2$, which measures the imbalance
between the 2-vertex types with positive ($\rho_{2+}$) and negative
($\rho_{2-}$) magnetization.

In Figs.~\ref{fig:eta-f}(a) and (b) we show experimental data for
$\rho_2=\rho_{2+}+\rho_{2-}$ and $m^2$ in the $\eta-\tau$
plane. Fig.~\ref{fig:eta-f}(c) is a plot of the data for $\rho_2$ and
$m^2$ as a function of $\eta/\tau$, for the 31 values of $\tau$. The
collapse of the data confirms that $\rho_2$ and $m^2$ are indeed
controlled by a single parameter, the relative fugacity
$d=e^{4\eta/\tau}$. Notice that at $\eta=0$, in particular,
$\rho_2\approx 1/3$ and $m^2\approx 0$, consistent with the 8 vertices
having the same density. Because the system size is finite, the phase
transitions are rounded to smooth crossovers. Altogether, the
experimental data in Figs.~\ref{fig:config}, \ref{fig:8-types},
and~\ref{fig:eta-f} are consistent with the theoretical phase diagram
presented in Fig.~\ref{fig:annealing}(c)

{\bf Outlook} --
The results above indicate that toric-code-like phases can be
programmed in quantum hardware with only one- and two-body
interactions. With the currently available hardware, with
$ZZ$-interactions and $X$-fields, we succeeded in observing the
classical 8-vertex model. In the annealing trajectory we traverse the
region of the quantum phase diagram where the $\mathbb{Z}_2$ quantum
spin liquid state resides, but the size of the many-body gap is too
small to measure. A $\mathbb Z_2$ quantum spin liquid with a sizable
gap can be implemented if $XX$-interactions and $Z$-fields are
available~\cite{CGY}. The current hardware already has the
latter. Extending the capability of the device to include
$XX$-interactions, together with the theoretical notion of
combinatorial gauge symmetry, would enable a realistic platform to
build programmable topological qubits.

{\bf Acknowledgements} --
The work by S.Z. and C.C. is supported by
the DOE grant No. DE-SC0019275. We also acknowledge DOE support
through the granted access to the D-Wave DW-2000Q device at LANL.



\bibliography{reference}

\pagebreak
\clearpage
\onecolumngrid
\appendix

\section*{Supplementary Information}

\subsection*{Symmetry and Ground State Degeneracy}

The gauge symmetry of the Hamiltonian in Eq.~(\ref{eq:Hamiltonian}) at $\eta=0$ is the result of the following property of the matrix $W$: 
\begin{align}
  L^{-1}\;W\;R = W
  \;,
  \label{eq:automorphism}
\end{align}
where $R$ and $L$ are monomial matrices~\cite{CGY}. $R$ and $L$ can be viewed as transforming the gauge and matter spins, respectively, on each vertex. For example,
the following pair satisfies Eq.~(\ref{eq:automorphism})
on \textit{each site}:
\begin{align}
L=&
\begin{pmatrix}
0 & +1 & 0 & 0 \\
+1 & 0 & 0 & 0 \\
0 & 0 & 0 & -1 \\
0 & 0 & -1 & 0 
\end{pmatrix}
&
R=&
\begin{pmatrix}
-1 & 0 & 0 & 0 \\
0 & -1 & 0 & 0 \\
0 & 0 & +1 & 0 \\
0 & 0 & 0 & +1 
\end{pmatrix}
\label{eq:LR}
\;.
\end{align}
The two key points are that the matrix transformation $R$ is diagonal with an even number of $-1$'s, and that $L$ is uniquely determined by $R$ as a result of the constraint Eq.~(\ref{eq:automorphism}). This symmetry holds for any transverse field $\Gamma$, i.e., throughout the entire annealing cycle. In the special case when the system is in the classical limit, at the end of the annealing cycle ($\Gamma=0$), the 8 degenerate ground states are shown in Table~\ref{tab:ground_states}. Each of these states has energy $-8J$.
\begin{table}[h]
    \begin{tabular}{|c||c  c  c  c|c  c  c  c|}
    \hline
     \text{Index} & $\sigma^z_1$ & $\sigma^z_2$ & $\sigma^z_3$ & $\sigma^z_4$ & $\mu^z_1$ & $\mu^z_2$ & $\mu^z_3$ & $\mu^z_4$ \\
     \hline
     \hline
     1 & -1 & -1 & 1 & 1 & 1 & 1 & -1 & -1\\
     2 & -1 & 1 & -1 & 1 & 1 & -1 & 1 & -1\\
     3 & -1 & 1 & 1 & -1 & 1 & -1 & -1 & 1\\
     4 & 1 & -1 & -1 & 1 & -1 & 1 & 1 & -1\\
     5 & 1 & -1 & 1 & -1 & -1 & 1 & -1 & 1\\
     6 & 1 & 1 & -1 & -1 & -1 & -1 & 1 & 1\\
     7 & 1 & 1 & 1 & 1 & 1 & 1 & 1 & 1\\
     8 & -1 & -1 & -1 & -1 & -1 & -1 & -1 & -1\\
    \hline
    \end{tabular}
\caption{The 8 degenerate ground states at $\eta=0$ and $\Gamma=0$ at each
    vertex in the lattice. The matter spins are tethered to the gauge
    spins. Note the non-trivial structure; when all gauge spins are
    up/down the matter spins are in the same direction as the gauge
    spins, but when only two gauge spins are up/down the matter spins
    are opposite.}
\label{tab:ground_states}
\end{table}

\subsection*{Defect Analysis}

\begin{figure}[!h]
\centering
\includegraphics[width=0.47\textwidth]{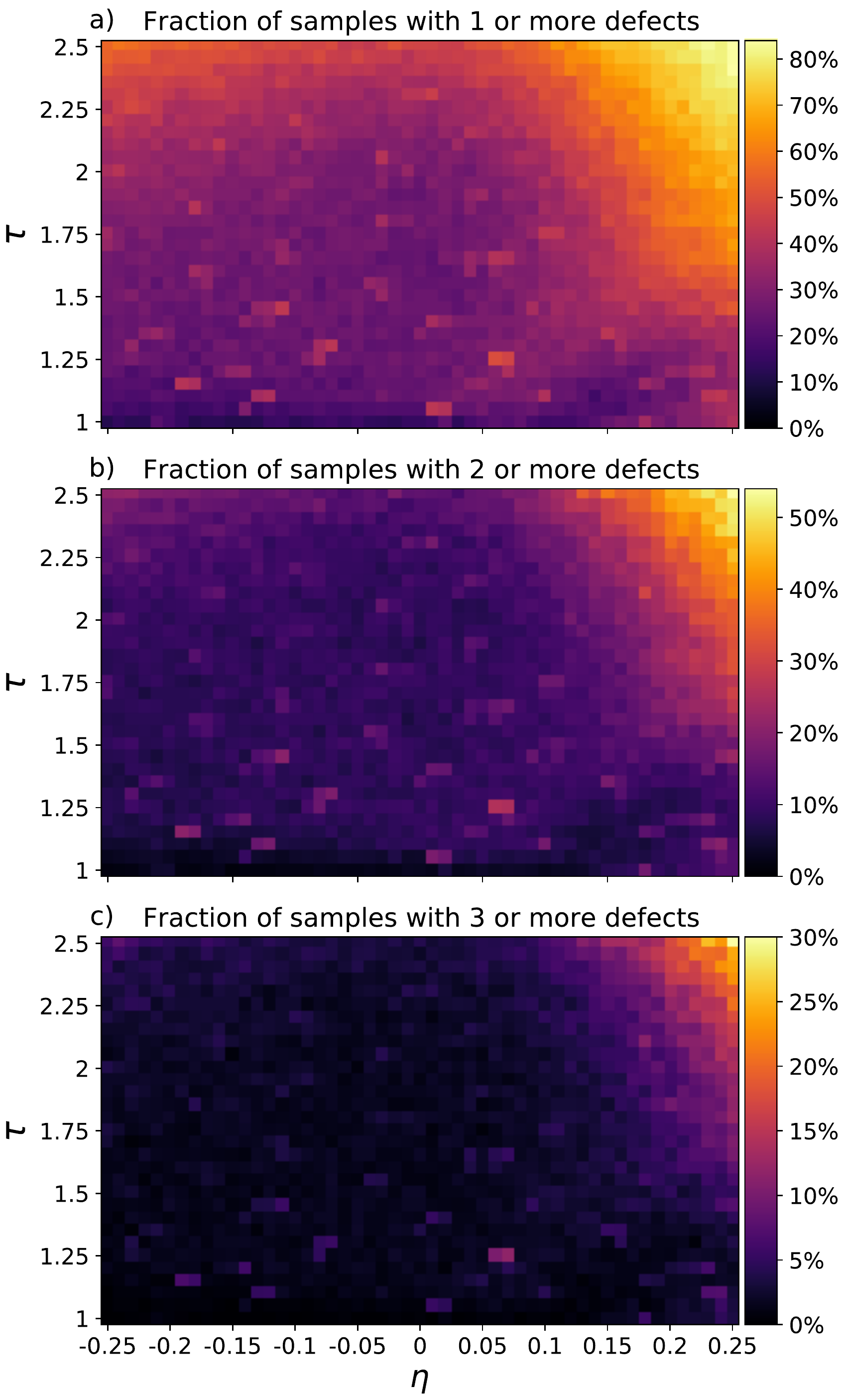}
\caption{Heat plot of the measured fraction of the 500 samples in each
  bin in the $\eta$ -- $\tau$ plane that have: (a) 1 or more defects;
  (b) 2 or more defects; and (c) 3 or more defects. The bins were
  constructed with $\tau$ varying from 1 to $2.5$ in intervals of
  $0.05$ and $\eta$ varying from $-0.25$ to $0.25$ in intervals of
  $0.01$.}
\label{fig:defects} 
\end{figure}

Two types of defects can occur in our embedding of the 8-vertex model:
(a) a negative parity star where the constraints imposed by the $J$
coupling and $W$ matrix are not respected (blue lines in
Fig.~\ref{fig:embedding} (a)); or (b) a broken gauge link where the
four physical qubits representing a single gauge qubit do not align
(red lines in Fig.~\ref{fig:embedding} (a)). The measurements were
taken at each bin across the entire phase digram in
Fig.~\ref{fig:eta-f}, with $\tau$ swept from 1 to $2.5$ in intervals
of $0.05$ and $\eta$ from $-0.25$ to $0.25$ in intervals of $0.01$.
For each bin, we collect 1000 samples and determine the fraction of
samples that have: 1 or more defects; 2 or more defects; and 3 or more
defects. The data for these three cases are shown as heat plots in
Fig.~\ref{fig:defects}. In our experiments, we applied spin reversal 
transformations on a set of spins to reduce the systematic biases in 
the D-Wave superconducting qubits~\cite{D-WaveDoc}. The total number 
of active stars and gauge links in one sample are 
$N_{\rm stars} = 114$ and $N_{\rm links} = 183$.

In the $\eta<0$ region, in the low temperature regime, we rarely
observe any defect. This is the region of the phase diagram where the
spin liquid is observed. At high temperatures, the $\eta<0$ region has
at most $58\%$ defects, and this fraction of samples with defects
mostly has a single defect. The $\eta>0$ region contains higher
fractions of defects, which we attribute to the formation of domain
walls in the ferromagnetic state. The highest defect fraction occurs
in the $\eta>0$ region at high temperatures, reaching as high as
$84\%$ of the samples. Other than in this corner of the plot, most 
samples rarely have more than 2 defects. 

If we aggregate all bins, as stated in the main text, there are in
total $31 \times 51 \times 500 = 1581000$ samples, out of which 1007751
($63.7\%$) have 0 defects, 374217 ($23.7\%$) have 1 defect, 146865
($9.3\%$) have 2 defects, and 38857 ($2.5\%$) have 3 defects.

\subsection*{Temperature Calibration}

Although the operating temperature in D-Wave is approximately 12 mK,
by scaling the coupling $J$ (with $K/J=1/2$ fixed) we effectively
control the ratios $T/J$ and $T/K$ in the experiments. The lowest
temperature corresponds to choosing the value $J_{\rm max}^{\rm API} =
1/2$, and the temperature is increased by decreasing the input value
$J^{\rm API} = J_{\rm max}^{\rm API} / \tau$, where $\tau$ increases
from 1 to $20$ in intervals of $1$ in the temperature calibration 
experiments. This factor $\tau$ can be expressed as
\begin{align}
  \tau = \frac{J^{\rm API}_{\rm max}}{J^{\rm API}} = \frac{T \, / \, J}{T / J_{\rm max}} = \frac{1}{T  /  J_{\rm max}} \frac{T}{J} = \alpha \frac{T}{J}
  \; ,
\end{align}
where $\alpha$ is a dimensionless ratio between the physical $J_{\rm
  max}$ (in units of mK) to the device temperature when $J^{\rm
  API}_{\rm max} = 1/2$.  The value of $\alpha$ can be pinpointed by
calibrating the experimental density of defects $\rho \, (\tau)$ with
the analytical probability of defects $p \, (T/J)$ calculated directly
from the model. We calibrate $\alpha$ with two variations of the
model where the theoretical probability of defects is easy to
calculate: (a) independent stars where $J^{\rm API}=J^{\rm API}_{\rm
  max}$ and $K^{\rm API}=0$ (or $J=J_{\rm max}$ and $K=0$); and
(b) independent gauge links where $J^{\rm API}=0$ and $K^{\rm
  API}=K^{\rm API}_{\rm max}=1$ (or $J=0$ and $K=K_{\rm
    max}$).

\begin{figure*}[h]
\centering
\includegraphics[width=1\textwidth]{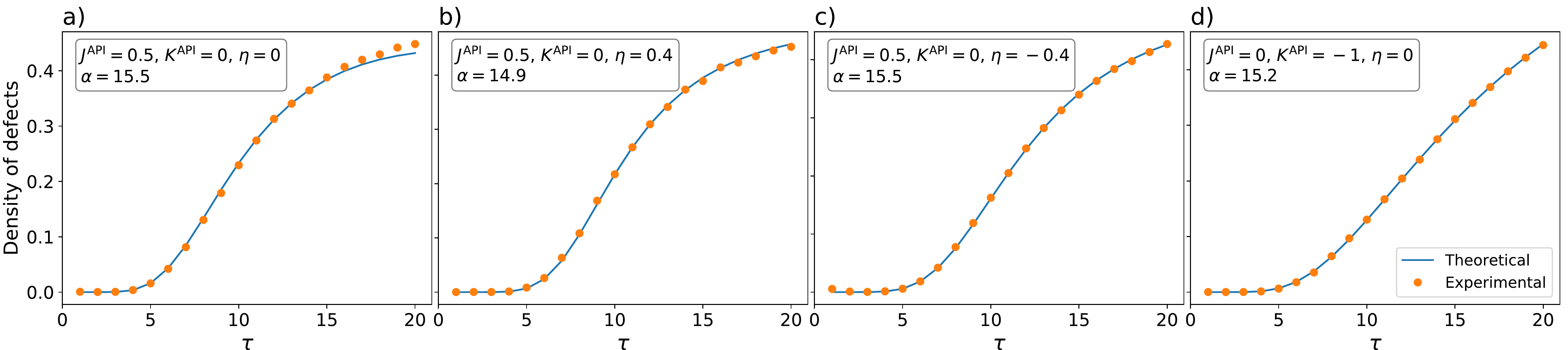}
\caption{Temperature calibration for two variations of the model. In
  (a), (b), and (c) we disconnect all the stars and form a
  configuration of $114$ independent stars by setting $K^{\rm
    API}=0$. The value of $\eta$ is set to 0 in (a), 0.4 in (b), and
  -0.4 in (c). In (d) we turn off the coupling $J$ by setting $J^{\rm
    API}=0$, and analyze the ferromagnetic chain of 4 physical qubits
  forming a gauge spin. We use the least squares to fit the analytical 
  probability of defects $p \, (T/J)$ to the experimental density of 
  defects $\rho \, (\tau)$, and extract the value of $\alpha$ that relates the
  experimental temperature control parameter $\tau$ to the physical
  ratio $T/J$.}
\label{fig:T-cali} 
\end{figure*}

The experimental density of defects $\rho \, (\tau)$ is defined as the
number of defective independent stars/links divided by the total
number of stars/links in a sample, averaged over 500 samples.  The
analytical probability of defects is given by $p \, (T / J) = Z_{\rm
  defects} \, / \, Z$, where $Z$ is the partition function, and
$Z_{\rm defects}$ is the partition function computed using only the
defective states. An independent star, built out of 4 gauge spins and
4 matter spins, has $2^8$ possible spin configurations with energies
determined by the Hamiltonian $H_{s} = 1/2 \, \sum_{a=1}^4
\sum_{i=1}^4 W_{ai} \, \sz_i \, \mz_a$, where $W_{ai}$ is the
interaction matrix between gauge spins $\sigma_i$ and matter spins
$\mu_a$ defined in Eq.~(\ref{eq:W}). From the single star (classical)
Hamiltonian $H_{s}$ we find the energy levels for all the $2^8$ spin
configurations. We then construct the partition function and compute
the $T / J$ dependence of the probability that a star has negative
parity given the four gauge spins, for $\eta = 0, \; 0.4, \;
\text{and} \; -0.4$. We compare the theoretical functions $p \, (T/J)$
to the experimental data $\rho \, (\tau)$ for $\tau$ varying from
1 to $20$, shown in Figs.~\ref{fig:T-cali} (a), (b), and (c). 
To illustrate this calculation, consider the case $\eta=0$. 
The full spectrum of the Hamiltonian in Eq.~(\ref{eq:Hamiltonian}) 
with $K = \Gamma = \widetilde\Gamma = 0$ is straightforward to compute, 
and is shown in Table~\ref{tab:spectrum}:
\begin{table}[h]
    \begin{tabular}{|c || c | c | c|}
    \hline
     Energy & \text{States} (P=+1) & \text{States} (P=-1) & \text{Total No. of States} \\
    \hline
    \hline
    8J & 8 & 0 & 8\\   
    4J & 32 & 64 & 96\\
    0J & 48 & 0 & 48\\   
    -4J & 32 & 64 & 96\\   
    -8J & 8 & 0 & 8\\
    \hline
    \end{tabular}
    \caption{The full spectrum and degeneracies of the Hamiltonian in 
             Eq.~(\ref{eq:Hamiltonian}) with $K=\Gamma=\widetilde\Gamma=0$ 
	     for the two different parities. The total number 
	     of states is $256=2^8$.}
    \label{tab:spectrum}
\end{table}
The corresponding density of negative parity stars is given by:
\begin{align}
    p(T/J)=\frac{8e^{4J/T}+8e^{-4J/T}}{e^{8J/T}+12e^{4J/T}+6+12e^{-4J/T}+e^{-8J/T}}~.
    \label{eq:probability}    
\end{align}
The other cases are calculated similarly. Using a least squares fit for 
the parameter $\alpha$, we find that $\alpha = 15.5$ for $\eta = 0$, $\alpha = 14.9$ 
for $\eta = 0.4$ and $\alpha = 15.5$ for $\eta = -0.4$.

A similar analysis can be done for the case of independent gauge
links. An independent gauge link consists of 4 physical spins
interacting via an Ising-type Hamiltonian $H_{l} = -1 \,
\sum_{i=1}^{3} s_i \, s_{i+1}$. Among $2^4$ states in total, the
defective states correspond to the configurations where the four spins
are not all aligned. We again construct the $T/J$ dependence of the
probability that a gauge link is defective, and find that $\rho \,
(\tau)$ fits $p \, (T/J)$ for $\alpha = 15.2$, shown in
Fig.~\ref{fig:T-cali} (d).

In conclusion, we fitted the defects density of the D-Wave
data to the theoretical defects probability calculated directly from
the model, and calibrated the value of $\alpha$ that
relates the experimental temperature control parameter $\tau$ to the
physical ratio $T/J$. The fact that $\alpha$ is close and approximately the same in all the cases above serves as another indication of robustness in our measurements.

\end{document}